\documentclass[3p,twocolumn]{elsarticle}
\usepackage{amsthm}
\usepackage{amsmath}
\usepackage{soul, color}
\usepackage{natbib} 
\usepackage{parskip} 
\usepackage{amsmath}
\usepackage{amsfonts}
\usepackage{float}
\usepackage{tabularx}
\usepackage{comment}
\usepackage{textcomp}
\usepackage{subfigure}
\usepackage{booktabs}
\usepackage{orcidlink}
\usepackage{tikz}
\usetikzlibrary{arrows.meta, positioning}
\usepackage{amsmath}
\usepackage{caption}

\usepackage{soul,color}

\usepackage{appendix}

\newcommand{\orcid}[1]{\href{https://orcid.org/#1}{\textcolor[HTML]{A6CE39}{\aiOrcid}}}

\journal{ArXiv}

\begin{document}
\begin{frontmatter}



\title{Modeling Membrane Degradation in PEM Electrolyzers with Physics-Informed Neural Networks}

\author[1]{Alejandro Polo-Molina\,\orcidlink{0000-0001-7051-2288}\corref{cor1}}
\ead{apolo@comillas.edu}

\author[1,2]{Jose Portela\,\orcidlink{0000-0002-7839-8982}}
\ead{jportela@comillas.edu}

\author[1]{Luis Alberto Herrero Rozas\,\orcidlink{0000-0002-0341-7333}}
\ead{lherrero@comillas.edu}

\author[3]{Román Cicero González\,\orcidlink{0000-0003-4418-6196}}
\ead{rcicero@innomerics.com}
\cortext[cor1]{Corresponding author}

\affiliation[1]{
  organization={Instituto de Investigación Tecnológica (IIT), Universidad Pontificia Comillas, Escuela Técnica Superior de Ingeniería ICAI},
  addressline={Alberto Aguilera 25},
  postcode={28015},
  city={Madrid},
  country={Spain}
}

\affiliation[2]{
  organization={Facultad de Ciencias Económicas y Empresariales, ICADE, Universidad Pontificia Comillas},
  addressline={Alberto Aguilera 23},
  postcode={28015},
  city={Madrid},
  country={Spain}
}

\affiliation[3]{
  organization={Innomerics},
  addressline={Segundo Mata, 4},
  postcode={28224},
  city={Pozuelo de Alarcón, Madrid},
  country={Spain}
}

\begin{abstract}
Proton exchange membrane (PEM) electrolyzers are pivotal for sustainable hydrogen production, yet their long-term performance is hindered by membrane degradation, which poses reliability and safety challenges. Therefore, accurate modeling of this degradation is essential for optimizing durability and performance. To address these concerns, traditional physics-based models have been developed, offering interpretability but requiring numerous parameters that are often difficult to measure and calibrate. Conversely, data-driven approaches, such as machine learning, offer flexibility but may lack physical consistency and generalizability. To address these limitations, this study presents the first application of Physics-Informed Neural Networks (PINNs) to model membrane degradation in PEM electrolyzers. The proposed PINN framework couples two ordinary differential equations, one modeling membrane thinning via a first-order degradation law and another governing the time evolution of the cell voltage under membrane degradation. Results demonstrate that the PINN accurately captures the long-term system's degradation dynamics while preserving physical interpretability with limited noisy data. Consequently, this work introduces a novel hybrid modeling approach for estimating and understanding membrane degradation mechanisms in PEM electrolyzers, offering a foundation for more robust predictive tools in electrochemical system diagnostics.
\end{abstract}




\end{frontmatter}

\section{Introduction}

Hydrogen is increasingly recognized as a pivotal element in the global energy transition, offering a clean, versatile, and sustainable energy carrier. Its importance stems from its ability to store and transport energy, helping to address one of the main challenges associated with renewable sources, their intermittency \cite{Barbir2005PEMSources, Sayed-Ahmed2024DynamicReview}. Hydrogen can be produced from various sources, including natural gas, coal, and water. Among these, water electrolysis powered by renewable energy, commonly referred to as \textit{green hydrogen}, offers a carbon-free alternative. Consequently, \textit{green hydrogen} is crucial for reducing greenhouse gas emissions and achieving climate goals, positioning it as a cornerstone of the future energy landscape \cite{Benmehel2024PEMReflections, Pivac2024ReductionElectrolysis}.

Within the various water electrolysis technologies, proton exchange membrane (PEM) electrolysis stands out as a promising option, as it can directly leverage the extensive research and technological advances made in PEM fuel cells (PEMFC) \cite{Liso2018ModellingTemperatures}. Compared to alkaline systems, PEM electrolyzers provide higher efficiency, better hydrogen purity, and can operate at higher current densities \cite{Schalenbach2013PressurizedCrossover, Falcao2020ABeginners}. Moreover, PEM electrolyzers offer a compact design and compatibility with renewable energy sources, making them well-suited for modern energy applications
\cite{Afshari2021PerformanceElectrolyzer,Li2024OptimizationPhenomenon,Arunachalam2024EfficientApproach}. Therefore, the use of PEM electrolyzers is expected to expand significantly in the coming years.

As PEM electrolyzers become a more attractive technology, developing accurate models is essential to support both performance optimization and durability enhancement \cite{Garcia-Valverde2012SimpleValidation}. 
A central challenge in this context is membrane degradation through membrane thinning, which results from chemical, mechanical and thermal stresses, significantly affecting the system’s long-term reliability and safety \cite{Chandesris2015MembraneDensity}.
Accurately modeling this phenomenon is key not only to extending device lifespan but also to mitigating safety risks. In particular, as the membrane thins, the likelihood of hydrogen and oxygen crossover increases. This crossover refers to the transportation of the gases through the electrolyte, which can lead to their mixing in the opposite electrode chamber. Such mixing poses a significant safety concern, as it increases the risk of forming explosive gas mixtures \cite{Trinke2016HydrogenConditions}. Therefore, detailed modeling of membrane degradation is vital to ensure both the sustained performance and secure operation of PEM electrolyzers.



To address these concerns, various physics-based models have been proposed to predict membrane degradation in PEM electrolyzers, relying on established electrochemical kinetics, transport phenomena, and thermodynamic principles \cite{Chandesris2015MembraneDensity, Frensch2019ImpactEmission, Sorace2022DevelopmentPhenomenon, Feng2017AStrategies}. These models offer interpretable insights, making them an established foundation for analyzing membrane degradation. However, they often lead to complex systems of equations that require numerical solutions, which can be computationally intensive and sensitive to the chosen discretization scheme \cite{Falcao2020ABeginners}. Moreover, their accuracy depends on a large set of material and kinetic parameters, such as the charge transfer coefficient or the degradation rate constant \cite{Sorace2022DevelopmentPhenomenon}. These parameters are difficult to measure directly and often vary with operation conditions making the models challenging to calibrate and limiting their predictive capability under real-world conditions \cite{Siegmund2021CrossingElectrolysis}.


To address the challenges of purely physics-based models, data-driven approaches, especially machine learning (ML) models, have become useful tools for predicting degradation in PEM electrolyzers \cite{Bahr2020ArtificialStacks, Hayatzadeh2024MachineCatalyst}. These approaches do not rely on explicitly defined physical equations. Instead, ML models learn system behavior directly from observed data, enabling flexible modeling even when the underlying mechanisms are complex or not fully understood. As a result, ML models are especially valuable in scenarios where parameter uncertainty, nonlinearity, or high-dimensional dynamics hinder traditional modeling. 

Despite the flexibility of the aforementioned ML approaches, conventional ML models often act as black boxes, lacking physical interpretability and extrapolating poorly outside the training domain \cite{Courtois2023CanDomingos}. Furthermore, their performance and reliability are intrinsically tied to the quantity and quality of available training data, which is often scarce in many physics-related projects. This has led to growing interest in hybrid approaches that integrate physics and data to combine the strengths of both paradigms. Among the hybrid approaches, physics-informed neural networks (PINNs) represent a significant advancement in the field of artificial neural networks (ANNs) by incorporating physical laws directly into the learning process \cite{Raissi2019Physics-informedEquations}. 
Unlike traditional ANNs, PINNs combine data-driven learning with the enforcement of governing equations, in the form of differential equations, as soft constraints in the loss function. This integration allows PINNs to improve generalizability and achieve high accuracy even with limited data \cite{Hu2024Physics-informedApplications, Moradi2023NovelPDEs}. Consequently, PINNs can substantially reduce the reliance on large datasets, making them a powerful tool for modeling complex systems where obtaining experimental data is costly.


One of the key advantages of PINNs is their ability to address both forward and inverse problems, offering unique solutions for a broad range of engineering challenges \cite{Raissi2019Physics-informedEquations}. In a forward problem, the goal is to predict the system's behavior based on known inputs, such as boundary conditions and parameters. On the other hand, inverse problems involve determining unknown parameters or model inputs based on observed data. Numerical approaches to inverse problems often require extensive computational resources and repeated simulations, which become prohibitively expensive in high-dimensional systems \cite{Chen2024Physics-InformedGeoengineering}. Moreover, PINNs exhibit strong robustness against noisy data, allowing for reliable inference of underlying physical parameters even under imperfect measurement conditions \cite{Raissi2019Physics-informedEquations}. These features make PINNs particularly well-suited for solving inverse problems in complex systems where data may be sparse, noisy, or costly to obtain \cite{Mishra2022EstimatesPDEs}.

The aforementioned advantages are especially relevant in systems like PEM electrolyzers, where critical degradation-related parameters are difficult to access experimentally and consequently poorly defined in the literature.
For instance, a key challenge arises from the absence of a well-defined kinetic constant for the overall membrane thinning mechanism \cite{Chandesris2015MembraneDensity, Frensch2019ImpactEmission}. This constant, which governs the degradation rate, remains poorly established in the literature, leading to substantial uncertainty in the predictive capacity of traditional models \cite{Chandesris2015MembraneDensity}. Consequently, the applicability of traditional models is limited when simulating long-term behavior or accounting for dynamic, aging-dependent conditions. 
 
Despite that some studies have highlighted the need for using PINN in the context of PEM electrolyzers, especially for determining difficult-to-specify parameters \cite{Bensmann2022AnDevelopment}, no study has yet explored their application to model the degradation through membrane thinning. However, PINNs have been successfully applied in other areas of PEM technologies, demonstrating their versatility and potential \cite{Zerrougui2025Physics-InformedElectrolysis, Chen2024MachineFramework}. 

For example, a recent study used PINNs to model and predict temperature fluctuations in PEM electrolysis \cite{Zerrougui2025Physics-InformedElectrolysis}. This approach outperformed traditional models, such as Long Short-Term Memory (LSTM), in terms of precision and robustness, even in the presence of sensor noise.
Additionally, in \cite{Chen2024MachineFramework} a knowledge-integrated machine learning framework is proposed to enhance overpotential prediction by embedding domain knowledge at multiple stages of the modeling process.
Although this work relates indirectly to membrane degradation, it focuses on estimating activation losses as a proxy rather than modeling the underlying physical degradation processes. In contrast, our work presents what is, to the best of our knowledge, the first PINN-based framework explicitly developed to model membrane degradation through membrane thinning in PEM electrolyzers. As mentioned before, this is a critical advancement, as membrane thinning is directly linked to increased gas crossover, which poses serious safety risks due to the formation of potentially explosive hydrogen–oxygen mixtures.
To do so, the proposed PINN framework aims to predict both the total cell voltage and the membrane thickness, which are critical factors influencing the performance and longevity of the electrolyzer. Consequently, the PINN is built upon two coupled ordinary differential equations (ODEs) that describe the system’s electrochemical dynamics and degradation behavior simultaneously. The first ODE governs the cell voltage, decomposed into reversible potential, activation, and ohmic overpotentials. This equation ensures that the electrochemical dynamics are consistent with established physical principles. The second ODE describes the thinning of the membrane due to chemical degradation, incorporating a first-order kinetic law to capture the degradation process.



This innovative application of PINNs represents a significant step forward in accurately modeling the degradation of PEM electrolyzers. It not only provides a tool to predict the system's long-term behavior given a limited set of data, but also formulates the problem as an inverse problem in which the degradation rate constant is inferred directly from observed data. By jointly estimating this poorly established kinetic parameter and the evolving physical state of the membrane, the proposed approach enables reliable long-term predictions without relying on ad hoc fitting or extensive degradation testing. This dual capability enhances the potential of PINNs as a data-efficient, physically consistent framework for both system identification and forecasting in PEM electrolyzers.


The structure of the paper is as follows: Section \ref{sec:PEM_Modelling} and \ref{sec:Mem_Modelling} present the numerical model describing the voltage evolution and the membrane degradation, respectively. Then, Section \ref{sec:PINN_Description} presents the theoretical background and formulation of the PINN approach. Section \ref{sec:experimental_setup} presents the methodology for the PINN training and the differential equations used as physics laws. Next, the results of the proposed approach are presented in Section \ref{sec:results_discussion}. Lastly, Section \ref{sec:conclusion} describes the conclusions obtained and the insights into future research directions. Moreover, the code of the proposed algorithm and the results are available at \href{https://github.com/alejandropolo/PEMElectrolyzerPINN}{\url{https://github.com/alejandropolo/PEMElectrolyzerPINN}}

\section{PEM Electrolyzer Modelling}\label{sec:PEM_Modelling}

Modeling of PEM electrolyzers is essential for optimizing their performance and understanding their behavior under different operating conditions. The PEM electrolyzer comprises an ion-exchange membrane, typically a perfluorinated sulfonic acid ionomer, placed between two electrodes. When supplied with electrical energy, it facilitates electrochemical water splitting, producing hydrogen and oxygen gases \cite{Garcia-Valverde2012SimpleValidation}. This process takes place within the electrochemical cell, where the following half-cell reactions occur
\begin{equation*}
\text{Anode:} \quad H_2O \rightarrow \frac{1}{2} O_2 + 2H^+ + 2e^-
\end{equation*}
\begin{equation*}
\text{Cathode:} \quad 2H^+ + 2e^- \rightarrow H_2.
\end{equation*}

Since the electrolysis of water is a non-spontaneous process, an external voltage must be applied to convert the electrical energy into chemical energy. Ideally, this voltage corresponds to the reversible potential (1.23 V) determined by the Gibbs free energy of the reaction \cite{Schalenbach2013PressurizedCrossover}. However, in practical operation, a higher cell voltage is required to overcome various internal losses \cite{Garcia-Valverde2012SimpleValidation, Marangio2009TheoreticalProduction}. These losses arise from reaction kinetics, ionic resistance within the membrane, and limitations in mass transport \cite{Liso2018ModellingTemperatures, Afshari2021PerformanceElectrolyzer, Marangio2009TheoreticalProduction}. As a result, the total cell voltage \( V \) includes contributions from different overpotentials and can be expressed as

\begin{equation*}
V = V_{\text{oc}} + V_{\text{act}} + V_{\text{ohm}},
\end{equation*}

where \( V_{\text{oc}} \) is the open-circuit voltage, determined by the minimum voltage required to initiate the electrochemical reaction in the absence of losses, \( V_{\text{act}} \) is the activation overpotential, and \( V_{\text{ohm}} \) is the ohmic overpotential.

Assuming ideal gas behaviour, the open-circuit voltage \( V_{\text{oc}} \) is calculated using the Nernst equation, which adjusts the standard reversible voltage for the actual temperature and pressure conditions of the system

\begin{equation*}
V_{\text{oc}} = E^0 + \frac{RT}{2F} \ln\left(\frac{p_{\mathrm{H}_2} \cdot \sqrt{p_{\mathrm{O}_2}}}{p_{\mathrm{H}_2O}}\right),
\end{equation*}

where \(E^0\) is the standard cell potential, \(R\) is the ideal gas constant, \(T\) is the temperature, and \(F\) is Faraday’s constant. This formulation assumed that only hydrogen and oxygen gases are present in their respective compartments, with hydrogen at the anode and oxygen at the cathode and that the partial pressure of water vapor, \(p_{\mathrm{H}_2O}\), is approximated by the saturation vapor pressure at the given operating temperature \cite{Sorace2022DevelopmentPhenomenon,Biaku2008AElectrolyzer}.

On the other hand, the activation overpotential $V_{\text{act}}$ reflects the kinetic barriers associated with the electrochemical reaction \cite{Liso2018ModellingTemperatures}. This effect is often modeled using the Butler–Volmer equation for both electrodes

\begin{equation*}
V_{\text{act}} = \frac{RT}{\alpha_{\text{an}}F} \ln\left(\frac{i}{i_{0,\text{an}}}\right) + \frac{RT}{\alpha_{\text{cat}}F} \ln\left(\frac{i}{i_{0,\text{cat}}}\right),
\end{equation*}

where \( i \) is the current density, \( i_{0,an} \) (resp. \( i_{0,cat} \)) is the exchange current density for the anode (resp. cathode), and \( \alpha \) are the respective charge transfer \cite{Liso2018ModellingTemperatures, Garcia-Valverde2012SimpleValidation}.

Finally, the ohmic overpotential \( V_{\text{ohm}} \) reflects the resistance to electrons' movement within the cell components, especially across the membrane \cite{Marangio2009TheoreticalProduction}. This loss can be described by a simple linear equation
\begin{equation*}
V_{\text{ohm}} = R_{\text{ohmic}} \cdot i,
\end{equation*}
where \( R_{\text{ohmic}} \) is the membrane’s ohmic resistance. The resistance depends on the membrane thickness \( t_{\text{mem}} \) and its ionic conductivity \( \sigma_{\text{mem}} \), and is given by
\begin{equation}\label{eq:ohmic_resistance}
R_{\text{ohmic}} = \frac{t_{\text{mem}}}{\sigma_{\text{mem}}}.
\end{equation}
Moreover, the conductivity \( \sigma_{\text{mem}} \) can be estimated empirically, as proposed in \cite{Carmo2013AElectrolysis}, following the equation below
\begin{equation*}
\sigma_{\text{mem}} = (0.005139 \cdot \lambda - 0.00326) \, \exp\left[1268 \, \left(\frac{1}{303} - \frac{1}{T}\right)\right],
\end{equation*}
where \( \lambda \) represents the membrane’s hydration level \cite{Sorace2022DevelopmentPhenomenon,Marangio2009TheoreticalProduction}.

\section{Membrane Degradation Modelling}\label{sec:Mem_Modelling}

Membrane degradation in PEM water electrolyzers is a critical factor affecting the longevity and performance of these systems. Despite this fact, available literature on this topic is relatively limited, especially when compared to the extensive research on PEMFCs \cite{Sorace2022DevelopmentPhenomenon}. However, the findings reported in existing studies converge toward similar conclusions regarding the dominant degradation mechanisms \cite{Chandesris2015MembraneDensity}, \cite{Frensch2019ImpactEmission}, \cite{Feng2017AStrategies}, \cite{Sorace2022DevelopmentPhenomenon}.

According to \cite{Chandesris2015MembraneDensity}, degradation is reported to occur due to various chemical, mechanical and thermal stresses that the membrane experiences during operation. Among the different factors that can contribute to degradation, membrane thinning is particularly critical \cite{Chandesris2015MembraneDensity,Sorace2022DevelopmentPhenomenon}. In particular, the electrolyzer gradual thinning generates concerns over system safety, long‐term durability and operational performance. Consequently, understanding and modeling membrane thinning is necessary for predicting the lifespan and ensuring the reliability of PEM electrolyzers.

One of the most commonly reported chemical degradation pathways in PEM electrolyzers involves a sequence of reactions triggered by gas crossover \cite{Chandesris2015MembraneDensity}, \cite{Frensch2019ImpactEmission}, \cite{Sorace2022DevelopmentPhenomenon}. As oxygen diffuses from the anode to the cathode side, it can lead to the formation of hydrogen peroxide $(\mathrm{H_2O_2})$, which subsequently breaks down into highly reactive radical species, such as hydroxyl $(\mathrm{HO^{\cdot}})$ or hydroperoxyl $(\mathrm{HOO^{\cdot}})$, via homolysis or Fenton's reaction \cite{Wallnofer-Ogris2024AResearch}. These free radicals can attack the membrane surface, reducing its thickness and causing fluoride ion release as a consequence of the erosion \cite{Wallnofer-Ogris2024AResearch}. 

Moreover, as the membrane becomes thinner, gas crossover increases, enhancing the rate of hydrogen peroxide and radical formation. This feedback loop leads to a non-linear, rather exponential,  decay in membrane thickness \cite{Chandesris2015MembraneDensity}. Such accelerated degradation not only compromises performance but also poses a safety risk due to the increased likelihood of forming explosive hydrogen–oxygen mixtures \cite{Kim2021AccurateMembranes}. Specifically, whenever the $\mathrm{H_2}$ concentration in the $\mathrm{O_2}$ exceeds the explosivity limit, typically around 4 of \% $\mathrm{H_2}$ in $\mathrm{O_2}$, an explosive mixture can form, posing significant safety risks \cite{Chandesris2015MembraneDensity, PatrickTrinke2021ExperimentalElectrolyzers}. These insights highlight the necessity of incorporating membrane thinning and fluoride release dynamics into any long-term performance and safety model for PEM electrolyzers.

As described in \cite{Chandesris2015MembraneDensity}, oxygen diffuses from the anode through the hydrated $\text{Nafion}^{\tiny{\textregistered}}$
membrane governed by the convective–diffusive balance  
\begin{equation*}
v_{\mathrm{H_2O}} \cdot \nabla c_i \;=\;\nabla \cdot \!D_i\nabla c_i,
\end{equation*}
where \(c_i\) is the local concentration of O\(_2\) (or H\(_2\)), \(D_i\) its diffusion coefficient, and \(v_{\mathrm{H_2O}}\) the electro‐osmotic drag water velocity. 

When the permeated oxygen reaches the cathode catalyst layer, it undergoes an oxygen reduction reaction (ORR)  
\begin{equation*}
\mathrm{O_2} + 2\,\mathrm{H}^+ + 2\,e^- \;\longrightarrow\; \mathrm{H_2O_2},
\end{equation*}
at a rate  
\begin{equation*}
R_1 \;=\; k_1\,c_{\mathrm{O_2}}\,c_{\mathrm{H}^+}^2,
\end{equation*}
where \(k_1\) is the kinetic constant of the reaction. Under low potential conditions, the ORR is assumed to proceed primarily through the hydrogen peroxide ($\mathrm{H_2O_2}$) formation pathway \cite{Chandesris2015MembraneDensity}. Hence, under this regime, water production recombination can be neglected \cite{Ruvinskiy2011UsingReaction}.


Once generated, the produced $\mathrm{H_2O_2}$ readily decomposes, either via direct reaction (homolysis of  $\mathrm{H_2O_2}$)
\begin{equation*}
\mathrm{H_2O_2} \longrightarrow 2\,\mathrm{HO}^\cdot
\end{equation*}
or through Fenton reaction with presence of metal ions  
\begin{equation*}
\mathrm{H_2O_2} + \mathrm{Fe}^{2+} \longrightarrow \mathrm{Fe}^{3+} + \mathrm{HO}^\cdot + \mathrm{HO}^-.
\end{equation*}
The resulting hydroxyl (\(\mathrm{HO}^\cdot\)) and hydroperoxyl (\(\mathrm{HOO}^\cdot\)) radicals attack the polymer backbone of the membrane, leading to cleavage of side chains, release of fluoride ions, and gradual thinning of the membrane surface. The overall dynamics of species concentrations in the cathode catalyst layer, including \(\mathrm{H_2O_2}\), \(\mathrm{HO}^\cdot\), \(\mathrm{HOO}^\cdot\), \(\mathrm{Fe}^{2+}\), and \(\mathrm{Fe}^{3+}\), are described by  
\begin{equation*}
\frac{d c_j}{d t} = \sum_i (-\text{sign}(\nu_{ji})\delta_{ji})\,v_i + \frac{j_j^{\mathrm{in}} - j_j^{\mathrm{out}}}{e_{\mathrm{cl}}},
\end{equation*}
where $\nu_{ji}$ and \(v_i\) are the stoichiometric coefficients and rates of the radical‐forming reactions, \(j_j^{\mathrm{in}}\) are the incoming fluxes and \(j_j^{\mathrm{out}}\) the convective fluxes carried by water, and \(e_{\mathrm{cl}}\) the catalyst‐layer thickness. Altogether, these coupled equations provide a framework for quantifying how oxygen crossover induces free‐radical attack, ultimately compromising membrane integrity and performance under prolonged operation.


Consequently, building on the previously described membrane attack mechanism, the chemical degradation of the membrane, and thus the evolution of its thickness \( t_{\mathrm{mem}}\), can be described using a simplified one-dimensional model. As mentioned before, the hydroxyl radicals (\(\mathrm{HO}^{\cdot}\)) initiate membrane decomposition by cutting the side chains of the $\text{Nafion}^{\tiny{\textregistered}}$ polymer from the backbone \cite{Chandesris2015MembraneDensity,Sorace2022DevelopmentPhenomenon}. This process further degrades in a self-propagating ``unzipping" reaction, releasing fluoride ions (as $\mathrm{HF}$) in the process. Therefore, the degradation reaction, following the pathway reported in \cite{Wong2014MacroscopicCells}, can be simplified as
\begin{equation*}
\mathrm{HO}^{\cdot} + \mathrm{R_fCF_2COOH} \longrightarrow \text{products} + n\mathrm{HF},
\end{equation*}
where the released HF represents the measurable fluoride ions in the effluent, serving as a direct indicator of chemical attack. Based on the work of \cite{Chandesris2015MembraneDensity}, the fluoride formation rate \(v_{\mathrm{fluor}}\) is related to the primary radical attack rate \(v_{5}\) as
\begin{equation*}
v_{\mathrm{fluor}} = 3.6 \cdot v_{5},
\end{equation*}
where $v_{5}$ is the reaction rate of the aforementioned chemical reaction and is given by
\begin{equation*}
v_{5} = k_{5} \cdot C_{\mathrm{HO}^{\cdot}} \cdot C_{\mathrm{mem}}.
\end{equation*}
Here, \(k_{5}\) is the kinetics constant, \(C_{\mathrm{HO}^{\cdot}}\) is the hydroxyl radical concentration, and \(C_{\mathrm{mem}}\) is the concentration of the $\text{Nafion}^{\tiny{\textregistered}}$ membrane, which depends on its density and equivalent weight.

To connect this chemical activity to a physical performance loss, the fluoride release rate (FRR) can be calculated as
\begin{equation*}
\mathrm{FRR} = v_{\mathrm{fluor}} \cdot MM_{\mathrm{F}} \cdot  t_{\mathrm{mem}} \cdot 3600 \frac{s}{h} \cdot 10^{-6} \frac{m^3}{cm^3} \cdot 10^6 \frac{\mu g}{g},
\end{equation*}
where \(MM_{\mathrm{F}} = 18.998\,\mathrm{g/mol}\) is the molar mass of fluoride and \( t_{\mathrm{mem}}\) is the membrane thickness in centimeters. The resulting FRR,  expressed in \(\mu\mathrm{g}/(\mathrm{h} \cdot \mathrm{cm}^2)\), represents the mass flow of fluoride ions over the area.

Hence, the FRR can then be linked to the reduction in membrane thickness through the thinning rate (TR) given by
\begin{equation*}
\mathrm{TR} = \frac{\mathrm{FRR}}{\rho_{\mathrm{Naf}} \cdot 0.82} \cdot 10^{-6} \frac{g}{\mu g} \; [\mathrm{cm/h}],
\end{equation*}
where \(\rho_{\mathrm{Naf}} = 2\,\mathrm{g/cm^3}\) is the $\text{Nafion}^{\tiny{\textregistered}}$ density and 0.82 accounts for the mass fraction of fluorine in the membrane. 

Therefore, the time evolution of the membrane thickness is governed by the following ordinary differential equation (ODE)
\begin{equation}\label{eq:TR_ODE}
\frac{d  t_{\mathrm{mem}}}{dt} = -\mathrm{TR}.
\end{equation}


Finally, although at first it might seem that reducing the thickness of the membrane could decrease the ohmic resistance (Eq.\,\ref{eq:ohmic_resistance}) and therefore increase the electrolyzer efficiency, as the membrane thins, the electrolyzer’s geometry changes and some mechanical properties of the membrane vary with time \cite{Chandesris2015MembraneDensity}. This is the case of the membrane's conductivity $\sigma$, that changes with the membrane thickness as follows
\begin{equation*}
    \sigma' = \left(\frac{ t_{\mathrm{mem}}}{ t_{\mathrm{mem}}^0}\right)^2 \cdot \sigma,
\end{equation*}
where \( t_{\mathrm{mem}}^0\) represents the initial membrane thickness \cite{Chandesris2015MembraneDensity}. Therefore, the Ohmic resistance (Eq.\,\ref{eq:ohmic_resistance}), and consequently the Ohmic overpotential, increases as the membrane thins, reducing the overall efficiency.

\section{Physics-Informed Neural Networks}\label{sec:PINN_Description}

Physics-Informed Neural Networks (PINNs) are an innovative class of Artificial Neural Networks (ANNs) that integrate the governing physical laws directly into the ANN training process \cite{Raissi2019Physics-informedEquations}. Unlike traditional ANNs, which primarily rely on data, PINNs leverage physical principles, such as conservation laws, as part of their loss function. This integration allows PINNs to achieve high accuracy with significantly less data and to provide better extrapolation capabilities \cite{Davini2021UsingNetworks}. Consequently, PINNs have demonstrated their effectiveness across a wide range of applications, including fluid dynamics \cite{Cai2021Physics-informedReview} or heat transfer \cite{Jalili2024Physics-informedFlows}.

The core idea behind PINNs is to embed the partial differential equations (PDEs) that describe the physical system into the learning process of the ANN. Consequently, during the training, the ANN not only minimizes the discrepancy between the predicted and observed data but also encourages the ANN's prediction to satisfy the underlying physical law. 

This dual-objective training strategy is implemented by adding to the loss function an additional term that penalizes deviations from the governing equations. As a result, the trained model is encouraged to respect the physics of the problem, leading to more accurate and reliable predictions.

Mathematically, consider a physical system described by a partial differential equation (PDE) of the form
\begin{equation*}
    \frac{\partial u}{\partial t} = \mathcal{D}(u(t,x)),
\end{equation*}
where \( u(t,x) \) represents the solution of the system, and \( \mathcal{D}[\cdot] \) is a differential operator encoding the relevant physical laws.

In the PINN framework, a neural network \( \hat{u}_\theta(t,x) \), parameterized by \( \boldsymbol{\theta} \), is trained to approximate the solution \( u(t,x) \) by minimizing a composite loss function defined as
\begin{equation*}
    \mathcal{L} = \lambda_{\text{data}} \cdot \mathcal{L}_{\text{data}}(\hat{u}_\theta, u^{\text{data}}) + \lambda_{\text{PDE}} \cdot \mathcal{L}_{\text{PDE}}(\hat{u}_\theta),
\end{equation*}
where \( \mathcal{L}_{\text{data}} \) measures the discrepancy between the predicted and observed data, and \( \mathcal{L}_{\text{PDE}} \) enforces consistency with the governing equation.

The data loss is typically computed using the mean squared error (MSE), defined as 
\begin{equation*}
    \mathcal{L}_{\text{data}} = \frac{1}{N_d} \sum_{i=1}^{N_d} \left| \hat{u}_\theta(\mathbf{z}_i) - u^{\text{data}}_i \right|^2,
\end{equation*}
where \( \mathbf{z}_i = (t_i, x_i) \) are the input coordinates of the \( N_d \) supervised data points. 

Similarly, the PDE residual loss is given by
\begin{equation*}
    \mathcal{L}_{\text{PDE}} = \frac{1}{N_r} \sum_{j=1}^{N_r} \left| \frac{\partial \hat{u}_\theta(\mathbf{z}_j)}{\partial t} - \mathcal{D}(\hat{u}_\theta(\mathbf{z}_j)) \right|^2,
\end{equation*}
where \( \mathbf{z}_j = (t_j, x_j) \) are collocation points sampled in the spatio-temporal domain. This term penalizes violations of the PDE and guides the model to produce physically consistent predictions. Consequently, by minimizing this joint loss, PINNs leverage both data and physical laws to learn robust, physically consistent representations of dynamical systems, even under limited or noisy observations.

Finally, it is important to note that the collocation points do not need to coincide with locations where observational data are available. This decoupling allows the PINN to enforce the governing physical laws throughout the entire spatio-temporal domain, even in regions where data are sparse or completely absent. As a result, the model can generalize better by leveraging physical priors, leading to improved robustness and extrapolation capabilities in real-world scenarios where data collection may be limited or expensive \cite{Raissi2019Physics-informedEquations, Davini2021UsingNetworks}.

\section{Experimental set up}\label{sec:experimental_setup}





\subsection{Data Collection}\label{sec:Data_Collection}


To train the PINN, we generated a synthetic dataset based on the numerical models described in Sections~\ref{sec:PEM_Modelling} and~\ref{sec:Mem_Modelling}. These models are grounded in well-established formulations that have been extensively validated against both experimental studies and operational data from PEM electrolyzers, providing a solid foundation for generating representative data of the inner behavior of PEM electrolyzers \cite{Liso2018ModellingTemperatures, Garcia-Valverde2012SimpleValidation,Chandesris2015MembraneDensity,Frensch2019ImpactEmission,Sorace2022DevelopmentPhenomenon}.

\begin{figure}[!t]
\centering
\includegraphics[width=\linewidth]{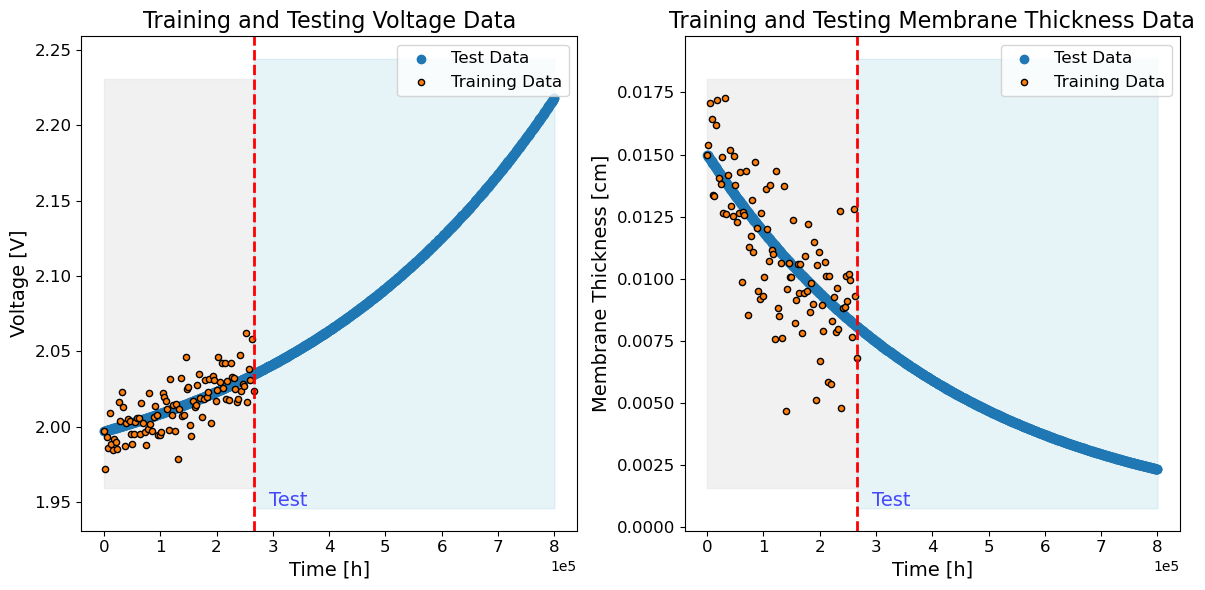}
\caption{Training and testing data used for model development and evaluation. \textbf{Left:} Voltage measurements over time, where orange circles represent the noisy training data and blue circles denote the test data used for testing interpolation and extrapolation beyond the training. \textbf{Right:} Membrane thickness measurements following the same color coding.}

\label{fig:training_test}
\end{figure}

Building on this foundation, we adopt a data generation strategy similar to \cite{Chen2024MachineFramework}, who relied on noisy synthetic data to address the limited availability of experimental measurements. Accordingly, we introduce random noise into the simulated data to emulate the variability and uncertainty characteristic of industrial operating conditions. Lastly, to evaluate the PINN’s ability to generalize beyond the training window and capture long-term degradation behavior, the network is trained only on the initial third of the noisy dataset and validated across the full degradation trajectory (Figure \ref{fig:training_test}). This strategy mimics real electrolyzer operation, where only a limited portion of the initial degradation trajectory is observable, while still aiming to accurately infer the long‐term degradation behavior.

Consequently, to generate the training and testing datasets, we simulated the time evolution of the cell voltage \(V(t)\) and the membrane thickness \( t_{\mathrm{mem}}(t)\) over a full operational cycle of \(8 \times 10^5\,\mathrm{h}\).
To reflect measurement uncertainty, we added Gaussian noise to each clean signal, using a standard deviation equal to that of the noise-free signal. Specifically, the noisy training signals for cell voltage \(\widetilde{V}(t)\) and membrane thickness \(\widetilde{t}_{\mathrm{mem}}(t)\) were defined as
\[
    \widetilde{V}(t) = V(t) + \epsilon_V(t), \quad \epsilon_V(t) \sim \mathcal{N}\big(0,\, \sigma_V^2\big),
\]
\[
    \widetilde{t}_{\mathrm{mem}}(t) =  t_{\mathrm{mem}}(t) + \epsilon_{\lambda}(t), \quad \epsilon_{\lambda}(t) \sim \mathcal{N}\big(0,\, \sigma_{t_{\mathrm{mem}}}^2\big),
\]
where \(\sigma_V\) and \(\sigma_{t_{\mathrm{mem}}}\) are the standard deviations of the noise-free signals \(V(t)\) and \( t_{\mathrm{mem}}(t)\), respectively.

Therefore, the proposed methodology allows us to closely approximate the constrained and noisy nature of real-world measurements encountered in practical operation. Moreover, by limiting the training data to early-stage degradation, we can rigorously test the PINN’s capability to extrapolate and accurately predict long-term system behavior.

\subsection{Governing Equations}\label{sec:Governing_Equations}

\begin{figure*}[!t]
\centering
\includegraphics[width=\linewidth]{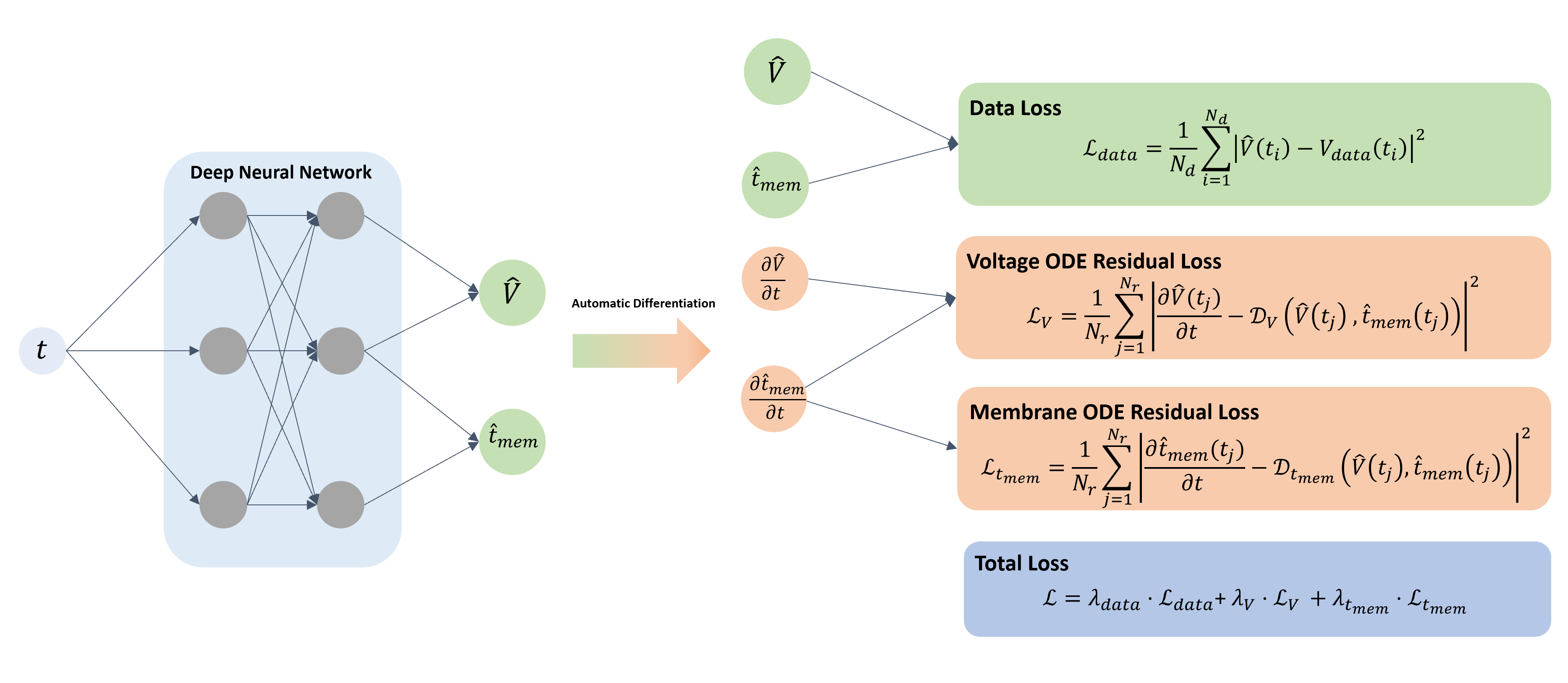}
\caption{PINN Framework for the Dual Estimation of Voltage and Membrane Thinning Evolution}
\label{fig:pinn_framework}
\end{figure*}

In order to capture both the electrochemical dynamics and the membrane‐degradation behavior of a PEM electrolyser, the proposed PINN incorporates two coupled ordinary differential equations as soft constraints. The first ODE enforces the cell voltage evolution, decomposed into open-circuit potential and the sum of activation and ohmic overpotentials. The second ODE governs membrane thinning due to the previously presented chemical attack. Therefore, following the PINN's methodology described in Section \ref{sec:PINN_Description}, both constraints are added to the loss term to ensure consistency with established physical laws. Hence, the PINN is trained to produce simultaneous approximations of both the cell voltage \(V(t)\) and the membrane thickness \( t_{\mathrm{mem}}(t)\) as functions of time, thereby learning the coupled electrochemical and degradation dynamics in one unified framework (Figure \ref{fig:pinn_framework}).

As described in Section \ref{sec:PEM_Modelling}, at any time \(t\), the total cell voltage is given by
\[
V(t)=V_{\mathrm{oc}}(t)+V_{\mathrm{act}}(t)+V_{\mathrm{ohm}}(t),
\]
where each term corresponds to the open-circuit, activation and ohmic potential, respectively. Moreover, we leverage Ohm's law to express the current density \(i(t)\) as the electric power divided by the product of active cell area \(A_\mathrm{cell}\) and voltage, i.e., \(i(t) = \frac{P}{A_\mathrm{cell} \cdot V(t)}\). This decision is based on the fact that, in the proposed PINN framework, the cell voltage \(V(t)\) is one of the outputs of the PINN, and that the physical loss is computed using the PINN's predictions.

Therefore, under the assumptions of constant temperature \(T\), constant cathode pressure $p_{\mathrm{cat}}$, and constant power input \(P\), the cell voltage equation can be expressed as 
\begin{multline}\label{eq:voltage_reduced} 
    V(t) = k_{1}^{V} + k_{2}^{V} \ln\!\left(\frac{P}{A_{\mathrm{cell}} \, V(t)}\right) \\
    + k_{3}^{V} \frac{1}{ t_{\mathrm{mem}}(t)}\frac{P}{A_{\mathrm{cell}} \, V(t)},
\end{multline}

where the constants \(k_1^V\), \(k_2^V\), and \(k_3^V\) are derived as follows
\begin{itemize}
    \item \(k_1^V = E^0 - \frac{RT}{2F} \ln\!\left(\frac{p_{\mathrm{H_2O}}}{p_{\mathrm{H_2}} \sqrt{p_{\mathrm{O_2}}}} \right) + \frac{RT}{\alpha F} \ln\!\left( \frac{1}{i_{0,\mathrm{an}} i_{0,\mathrm{cat}}} \right)\)
    
    \item \(k_2^V = \frac{2RT}{\alpha F}\)
    
    \item \(k_3^V = \frac{(t_{\mathrm{mem}}^0)^2}{\sigma}\)
\end{itemize}


Hence, in order to transform the cell voltage equation \eqref{eq:voltage_reduced} into an ODE suitable for training the PINN, we differentiate both sides with respect to time \(t\), yielding the following ODE that will be used as part of the physics-informed loss function
\begin{multline}\label{eq:Voltage_Ode_evolution}
\frac{dV}{dt} = -k_2^V \frac{1}{V(t)} \frac{dV}{dt} 
+ k_3^V \, \frac{P}{A_{\mathrm{cell}}} \\
   \cdot\left( - \frac{1}{V(t)  t_{\mathrm{mem}}^2(t)} \frac{d t_{\mathrm{mem}}}{dt} 
      - \frac{1}{ t_{\mathrm{mem}}(t) V^2(t)} \frac{dV}{dt} \right).
\end{multline}



On the other hand, membrane thinning is described by the following first‐order kinetics law (Eq. \eqref{eq:TR_ODE}) given by  
\begin{equation}\label{eq:mem_thinning_ode}
    \frac{d\, t_{\mathrm{mem}}}{dt}=-\frac{3.6 \cdot k_{5} \cdot C_{\mathrm{mem}}\cdot \mathrm{MM}_{\mathrm{F}} \cdot 3600 }{ 0.82\cdot\rho_{\mathrm{Naf}} \cdot 10^{6}}\cdot c_{\mathrm{HO}}\cdot t_{\mathrm{mem}}.
\end{equation}
As it can be observed, this degradation pathway is intrinsically linked to the presence of hydroxyl radicals \((\mathrm{HO}^{\cdot})\), whose concentration directly affects the thinning rate. To model this mechanism, and following the assumptions proposed in \cite{Sorace2022DevelopmentPhenomenon}, the framework neglects the presence of transition metal ions such as Fe\(^{2+}\) and Fe\(^{3+}\), which would otherwise catalyze radical formation via Fenton-type reactions. Consequently, the generation of hydroxyl radicals is attributed solely to the decomposition of hydrogen peroxide (\(\mathrm{H_2O_2}\)). Furthermore, the model assumes steady-state radical concentrations. A detailed list of the specific reactions involved in this mechanism is provided in the Appendix \ref{app:radical_reactions}.

Under the assumptions of steady-state operation and no metal ion presence, the temporal evolution of the involved concentrations is governed by the model proposed in \cite{Sorace2022DevelopmentPhenomenon}
\begin{subequations}\label{eq:ode_system}
\begin{align}
c_{\mathrm{HO}}(t)
=\frac{v_{\mathrm{H_2O}}(V(t))}{e_{\mathrm{cl}}\,k_{3}}
  -\frac{k_{2}}{k_{3}}
  -\frac{v_{1}}{k_{3}\,c_{\mathrm{H_2O_2}}(t)}, \notag 
\\[0.5em]
A(t)\,c_{\mathrm{H_2O_2}}(t)^2
+B(t)\,c_{\mathrm{H_2O_2}}(t)
+C(t) = 0, \notag
\end{align}
\end{subequations}
where \(v_{\mathrm{H_2O}}(V(t))\) is the water velocity, \(e_{\mathrm{cl}}\) is the thickness of the cathode catalyst layer and $A(t), B(t)$ and $C(t)$ are the coefficients of the quadratic equation given by
\begin{subequations}
\begin{align*}
A(t) &= -3k_{2} + \frac{v_{\mathrm{H_2O}}(V(t))}{e_{\mathrm{cl}}}, \\[0.5em]
B(t) &= \left(k_{4}\,c_{\mathrm{O_2}} + k_{5}\,C_{\mathrm{mem}} - \frac{v_{\mathrm{H_2O}}(V(t))}{e_{\mathrm{cl}}}\right)
       \\
     &\quad\cdot\left(\frac{v_{\mathrm{H_2O}}(V(t))}{e_{\mathrm{cl}}\,k_{3}}\right) \notag - v_{1}
      \\
     &\quad -\left(k_{4}\,c_{\mathrm{O_2}} + k_{5}\,C_{\mathrm{mem}} - \frac{v_{\mathrm{H_2O}}(V(t))}{e_{\mathrm{cl}}}\right) \cdot
       \left(\frac{k_{2}}{k_{3}}\right), \\[0.5em]
C(t) &= -\left(k_{4}\,c_{\mathrm{O_2}} + k_{5}\,C_{\mathrm{mem}} - \frac{v_{\mathrm{H_2O}}(V(t))}{e_{\mathrm{cl}}}\right)
        \cdot\left(\frac{v_{1}}{k_{3}}\right),
\end{align*}
\end{subequations}
with \(v_{1}\) the hydrogen peroxide formation rate, \(C_{\mathrm{mem}}\) the molar concentration of Nafion, and \(c_{\mathrm{O_2}}\) the local oxygen concentration in the cathode catalyst layer. The values of all parameters used in this formulation, along with further clarification of their definitions and units, can be found in Appendix~\ref{app:constants}.

It is important to note that, although the concentrations of \(\mathrm{HO}^{\cdot}\) and \(\mathrm{H_2O_2}\) are formally assumed to be in a steady state, their values still exhibit a parametric dependence on time through the cell voltage \(V(t)\) \cite{Sorace2022DevelopmentPhenomenon}. This dependence arises because the concentrations are functions of \(V(t)\), which itself evolves over time due to membrane degradation, as described in Eq.~\eqref{eq:Voltage_Ode_evolution}. Consequently, although the concentrations are not governed by explicit time dynamics in the equations, they inherit a time-varying behavior through their dependence on \(V(t)\).

Finally, let \(\widehat V(t)\) and \(\widehat{t}_{\mathrm{mem}}(t)\) denote the PINN’s predictions at time $t$ of the voltage and the membrane thickness, respectively. Then,  considering 
\begin{multline*}
    \mathcal{D}_V(V, t_{\mathrm{mem}}) = \Biggl(-k_2^V \frac{1}{V} \frac{dV}{dt} 
+ k_3^V \, \frac{P}{A_{\mathrm{cell}}} \\
  \cdot\left( - \frac{1}{V\,  t_{\mathrm{mem}}^2} \frac{d t_{\mathrm{mem}}}{dt} 
      - \frac{1}{ t_{\mathrm{mem}}\, V^2} \frac{dV}{dt} \right)\Biggl). 
\end{multline*}
the differential operator associated with the voltage's evolution residual loss and 
\begin{multline*}
    \mathcal{D}_{t_{\mathrm{mem}}}( V, t_{\mathrm{mem}}) = \\
-    \frac{3.6 \cdot k_{5} \cdot C_{\mathrm{mem}}\cdot \mathrm{MM}_{\mathrm{F}} \cdot 3600 }{\, 0.82 \cdot\rho_{\mathrm{Naf}} \cdot 10^{6}}\cdot c_{\mathrm{HO}}(V)\cdot t_{\mathrm{mem}},
\end{multline*}
the differential operator of the degradation law governing the membrane thinning, then the total loss function of the proposed PINN can be expressed as 
\begin{multline*}
\mathcal{L}
= \frac{1}{N}\sum_{i=1}^{N_d} \bigl|\widehat V(t_i)-V_{\mathrm{data}}(t_i)\bigr|^2
+ \\ \lambda_V \,\frac{1}{N_r}\sum_{j=1}^{N_r} \Bigg|\frac{\partial\,\widehat{V}(t_j)}{\partial \,t}-\mathcal{D}_V\left(\widehat{V}(t_j),\widehat{t}_{\mathrm{mem}}(t_j)\right)\Bigg|^2 
+ \\  \lambda_{t_{\mathrm{mem}}}\,\frac{1}{N_r}\sum_{j=1}^{N_r} \Bigg|\frac{\partial\,\widehat{t}_{\mathrm{mem}}(t_j)}{\partial\, t}-\mathcal{D}_{t_{\mathrm{mem}}}\left(\widehat{V}(t_j),\widehat{t}_{\mathrm{mem}}(t_j)\right)\Bigg|^2 
\end{multline*}
Besides, initial conditions of the voltage \(V(0)=V_0\) and the membrane thickness \( t_{\mathrm{mem}}(0)=t_{\mathrm{mem}}^0\) are enforced by boundary loss, ensuring a well‐posed initial‐value problem.

\subsection{Setup for the Joint Forecasting and Parameter Inference Using PINNs}

As previously discussed, the kinetic rate constant governing the membrane thinning reaction in PEM electrolyzers remains poorly characterized. To date, there is no widely accepted or precisely determined value for this parameter, despite its critical influence on membrane lifetime \cite{Chandesris2015MembraneDensity,Frensch2019ImpactEmission}. Therefore, by treating \(k_{5}\) as an additional trainable parameter within the PINN, we can leverage the noisy, partial observations of \(V(t)\) and \(t_{\mathrm{mem}}(t)\) not only to forecast future behavior but also to infer this key degradation constant in situ.  This dual capability is of great practical importance as it enables real‐time calibration of degradation models and provides long‐term voltage and membrane thinning predictions.


Therefore, using the proposed physics loss associated with the membrane thinning (Eq. \ref{eq:mem_thinning_ode}), we let \(\widehat{k}_{5}\) be a trainable scalar that is adjusted through training to approximate the real value. Consequently, during the PINN's training, gradients flow through this parameter just as they do through the network weights, minimizing the combined data‐mismatch and physics‐residual losses.

To guide this inference and improve convergence, we incorporate prior knowledge from the literature that suggests the kinetic rate constant \(k_{5}\) is expected to be on the order of \(10^{3}\, \big[\frac{\mathrm{m}^{3}}{\mathrm{mol}\,\mathrm{s}}\big]\) \cite{Chandesris2015MembraneDensity}. In practice, this is implemented by scaling the trainable parameter \(\widehat{k}_{5}\) with a fixed factor of \(10^{3}\) during the calculation of the physics-informed loss residual. This scaling ensures that \(\widehat{k}_{5}\) remains numerically well-conditioned within the optimization process and reflects realistic magnitudes encountered in PEM electrolyzer degradation kinetics. 

Consequently, the corresponding ODE differential operator used in the physics-informed loss is then defined as
\begin{multline}\label{eq:mem_thinning_ode_dimensionless}
    \mathcal{D}_{t_{\mathrm{mem}}}(\widehat{V}, \widehat{t}_{\mathrm{mem}};\, \widehat{k}_{5}) = \\
    \frac{3.6 \cdot \widehat{k}_{5} \cdot 10^{3} \cdot C_{\mathrm{mem}} \cdot \mathrm{MM}_{\mathrm{F}} \cdot 3600 }{0.82 \cdot \rho_{\mathrm{Naf}} \cdot 10^{6}} \cdot c_{\mathrm{HO}}(\widehat{V}) \cdot \widehat{t}_{\mathrm{mem}},
\end{multline}
where \(\widehat{k}_{5}\) is the dimensionless trainable parameter inferred by the PINN.



Moreover, to train the PINN, we have generated a synthetic dataset, following the methodology presented in Section~\ref{sec:Data_Collection}, using operating conditions of \(40\,^\circ\mathrm{C}\) temperature, \(30\,\mathrm{bar}\) cathode pressure, and a constant power load of \(500\,\mathrm{W}\). In addition, the initial membrane thickness and the cell active area are set to 0.0175 and \(680\,\mathrm{cm}^2\) respectively. From the resulting trajectory, only the first third of the time window is used for training. Within this subset, 100 equally spaced data points are sampled over a total simulated duration of \(8 \times 10^5\,\mathrm{h}\). To evaluate the model’s generalization and long-term forecasting capabilities, 1000 uniformly distributed data points spanning the entire operational cycle are used for testing (Figure~\ref{fig:training_test}).

\begin{figure*}\
\centering     
\subfigure[PINN prediction]{%
    \includegraphics[width=70mm]{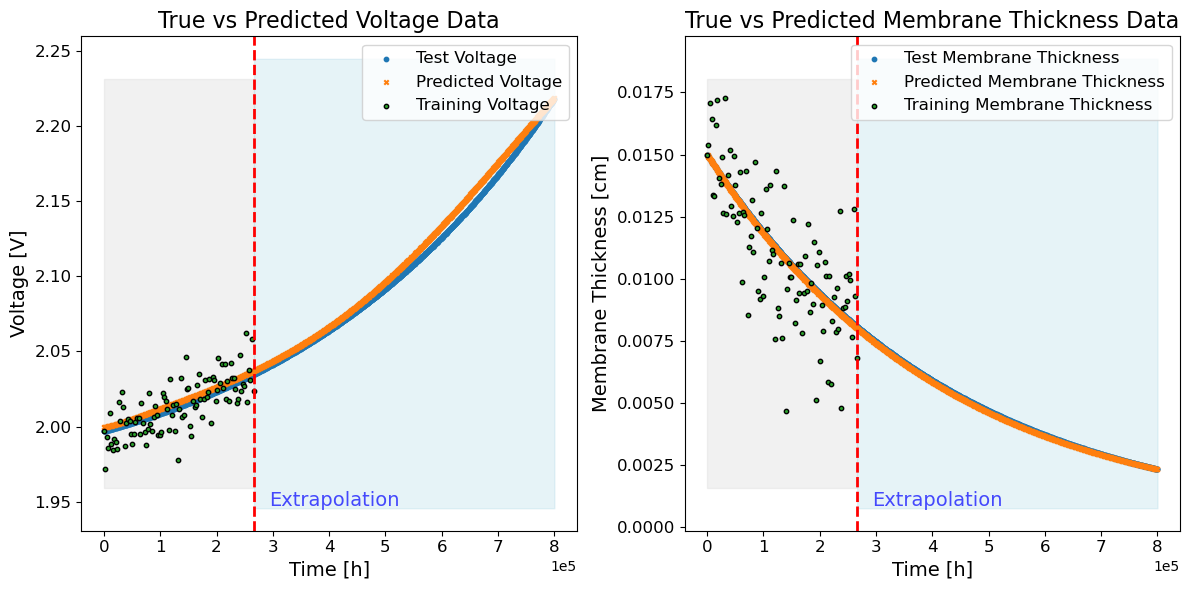}
    \label{fig:PINN_Training_a}
}
\hspace{7mm}
\subfigure[ANN prediction]{%
    \includegraphics[width=70mm]{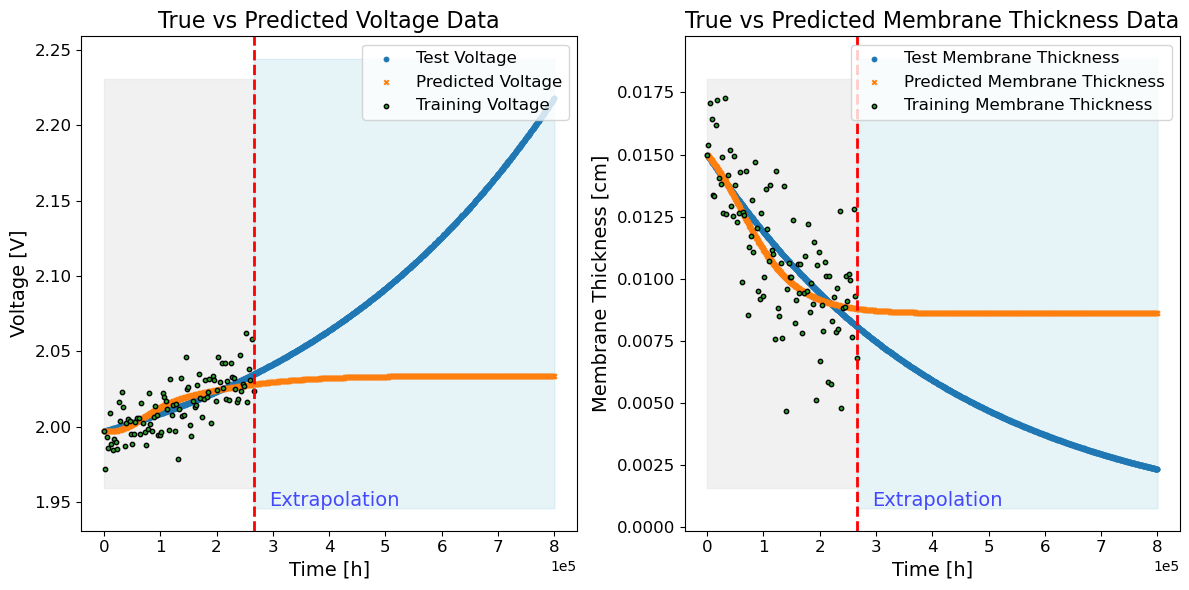}
    \label{fig:PINN_Training_b}
}
\caption{Comparison of prediction performance on voltage and membrane degradation data between the PINN and a conventional ANN. \textbf{Left}: True, predicted, and training values of cell voltage and membrane thickness over operation time using the PINN. \textbf{Right}: True, predicted, and training values of cell voltage and membrane thickness over operation time using the conventional ANN.}

\label{fig:PINN_Training}
\end{figure*}

The architecture employed for this task consists of a feedforward neural network with two hidden layers, the first one with 10 neurons and the second hidden layer with 5 neurons, both using sigmoid activations. Besides, as already mentioned, the trained neural network has two outputs to predict both membrane thickness and voltage over time. To train the PINN, optimization is performed using the Adam algorithm with a learning rate of \(0.005\) for a maximum of 7,500 epochs.

\section{Results and Discussion}\label{sec:results_discussion}

This section presents and analyzes the performance of the proposed PINN-based framework on the degradation data, focusing on its ability to infer hidden parameters and generalize beyond the training window. 

As observed in Figure \ref{fig:PINN_Training_a}, despite the limited amount of data and the presence of noise, the PINN successfully converges to a physically consistent solution. Notably, the inferred degradation constant \( \widehat{k}_{5} \) is initialized at zero, reflecting a complete lack of prior knowledge apart from the expected order of magnitude (Eq. \ref{eq:mem_thinning_ode_dimensionless}). Throughout the training process, however, the PINN progressively refines this estimate and ultimately converges to \( \widehat{k}_{5} = 1.01 \), which is remarkably close to the true normalized value of \( 1.0 \) used in the synthetic generator. This result demonstrates the model’s capacity to infer latent degradation kinetics from sparse and noisy observations. At the same time, the PINN produces stable and accurate long-term predictions for both \( V(t) \) and \( t_{\mathrm{mem}}(t) \) across the remaining 66\% of the cycle, highlighting the robustness of physics-informed learning under conditions of limited data and uncertain system dynamics. 
Quantitatively, the model achieves a training Root Mean Squared Error (RMSE) of \(0.0126\) for voltage prediction and \(0.00192\) for membrane thickness, and a significantly lower test RMSE of \(0.0047\) and \(0.000061\), respectively. This reduction in test error by more than an order of magnitude across both voltage and membrane predictions suggests that the embedded physics not only guides the network toward physically plausible solutions but also acts as an effective regularizer, preventing overfitting despite the limited and noisy training data. These values indicate that the model is capable of predicting voltage and membrane thinning with high accuracy, even in the presence of noise and limited observations.

To further evaluate the benefits of the physics-informed approach, we trained a conventional ANN using the same architecture, hyperparameters, and data conditions, but without incorporating the physics-based residual loss (Figure \ref{fig:PINN_Training_b}). While both models achieved similar training RMSEs, \(0.0127\) for voltage and \(0.00185\) for membrane thickness in the ANN, versus \(0.0126\) and \(0.00192\) in the PINN, their generalization performances diverged significantly. The ANN's test RMSEs reached \(0.0761\) for voltage and \(0.003532\) for membrane thickness, which are more than an order of magnitude higher than those obtained by the PINN (\(0.0047\) and \(0.000061\), respectively). This contrast highlights the critical role of embedded physics in enhancing generalization. While the ANN overfits to the training data and fails to extrapolate physically consistent degradation behavior, the PINN leverages domain knowledge to produce robust and interpretable long-term predictions.


\begin{table*}[t] 
\caption{Comparison of PINN and ANN performance for membrane degradation modeling under identical training conditions. RMSE values are reported separately for voltage and membrane predictions.}
\centering
\renewcommand{\arraystretch}{1.5}  
\begin{tabular}{ | >{\centering\arraybackslash}m{20em} | >{\centering\arraybackslash}m{4cm} | >{\centering\arraybackslash}m{4cm} | }
\hline
\textbf{Metric} & \textbf{Conventional ANN} & \textbf{PINN} \\
\hline 
Training RMSE (Voltage) [V]   & \(0.0127\) & \(0.0126\) \\
\hline
Testing RMSE (Voltage) [V]    & \(0.0761\) & \(0.0047\) \\
\hline
Training RMSE (Membrane) [cm]  & \(0.00185\) & \(0.00192\) \\
\hline
Testing RMSE (Membrane) [cm]  & \(0.00353\) & \(0.000061\) \\
\hline
\end{tabular}
\label{tab:comparison_pinn_ann}
\end{table*}


Therefore, this case study illustrates how the PINN framework can be deployed in the field to infer unseen degradation rates and to deliver reliable remaining‐life predictions under realistic noise and data‐scarcity conditions.  

\section{Conclusion}\label{sec:conclusion}

This study presents the first application of Physics-Informed Neural Networks (PINNs) to model membrane degradation through membrane thinning in Proton Exchange Membrane (PEM) electrolyzers. By embedding fundamental electrochemical and degradation dynamics into the neural network architecture, the proposed framework enables simultaneous long-term prediction of the cell voltage and membrane thickness, as well as inference of an otherwise difficult-to-measure kinetic degradation constant determining the degradation rate. 

Our results demonstrate that the PINN accurately learns the long-term degradation dynamics and the latent system parameters from sparse, noisy data, achieving high generalization performance and outperforming conventional data-driven approaches by several orders of magnitude in prediction accuracy. In particular, the trained PINN inferred the normalized degradation rate constant with high accuracy, despite being initialized with no prior knowledge. Moreover, the model showed strong robustness under realistic constraints, successfully extrapolating long-term degradation behavior from limited early-stage measurements.

Altogether, these findings highlight the potential of physics-informed learning for advancing the predictive modeling of PEM electrolyzers, especially in contexts where data are scarce and experimental characterization is challenging. By integrating physical knowledge into the learning process, PINNs offer a scalable and interpretable alternative to purely empirical or mechanistic models, facilitating real-time calibration, online monitoring and digital twin implementations.


As a direction for future work, the membrane‐thinning model could be enriched by incorporating explicit dependencies on operating conditions such as temperature, pressure, and humidity. Additionally, including Fenton-type reactions could increase the precision of the chemical degradation mechanism. Therefore, by embedding these additional physics‐informed terms into the PINN loss, the model could not only improve predictive accuracy but also uncover hidden relationships between operational variables and membrane degradation, ultimately guiding the optimization of electrolyzer performance and durability.


\section*{CRediT authorship contribution statement}

\textbf{Alejandro Polo}: Writing – review \& editing, Writing – original
draft, Visualization, Validation, Software, Methodology, Investigation,
Formal analysis, Data curation, Conceptualization. \textbf{Jose Portela}: Writing – review \& editing, Validation, Supervision, Resources, Project administration, Methodology, Funding acquisition. \textbf{Luis Alberto Herrero Rozas}: Writing – review \& editing, Validation, Supervision, Methodology. \textbf{Román Cicero González}: Writing – review \& editing, Validation, Supervision, Resources, Project administration, Methodology, Funding acquisition

\section*{Declaration of competing interest}
The authors declare that they have no known competing financial interests or personal relationships that could have appeared to influence the work reported in this paper.

\section*{Acknowledgments}
This research was supported by funding from CDTI, with Grant Number MIG-20221006 associated with the ATMOSPHERE Project.

\appendices
\section{Model Constants}
\label{app:constants}

This appendix provides a comprehensive summary of the physical and electrochemical constants used throughout the modeling framework. Unless otherwise stated, the values are taken the literature sources considered in Section \ref{sec:PEM_Modelling} and \ref{sec:Mem_Modelling}. Table~\ref{tab:constants} lists all the constants along with their respective symbols, descriptions, and numerical values.

\begin{table}[h]
\scriptsize
\centering
\caption{Summary of physical and electrochemical constants used in the model.}
\label{tab:constants}
\begin{tabular}{@{\hskip 1pt}l@{\hskip 2pt}p{2.0cm}@{\hskip 2pt}c@{\hskip 2pt}c@{\hskip 1pt}}
\toprule
\textbf{Symbol} & \textbf{Description} & \textbf{Value} & \textbf{Ref.} \\
\midrule
\( R \) & Ideal gas constant & \(8.314\,\mathrm{J/mol/K}\) & - \\
\( F \) & Faraday’s constant & \(96485\,\mathrm{C/mol}\) & - \\
\( A_{\mathrm{cell}} \) & Cell active area & \(680\,\mathrm{cm}^2\) & \cite{Mayyas2019ManufacturingElectrolyzers} \\
\( \alpha_{\mathrm{an}} \) & Anode transfer coefficient & \(0.5\) & \cite{Colbertaldo2017Zero-dimensionalElectrolyzers} \\
\( \alpha_{\mathrm{cat}} \) & Cathode transfer coefficient & \(0.5\) & \cite{Colbertaldo2017Zero-dimensionalElectrolyzers} \\
\( i_{\ell} \) & Limiting current density & \(6\,\mathrm{A/cm}^2\) & \cite{Garcia-Valverde2012SimpleValidation} \\
\( \lambda \) & Membrane hydration param. & \(20\) & \cite{Awasthi2011DynamicProduction} \\
\( i_{0,\mathrm{an}} \) & Anode exchange current dens. & \(2.3 \times 10^{-7}\,\mathrm{A/cm}^2\) & \cite{Liso2018ModellingTemperatures} \\
\( i_{0,\mathrm{cat}} \) & Cathode exchange current dens. & \(1.0 \times 10^{-3}\,\mathrm{A/cm}^2\) & \cite{Liso2018ModellingTemperatures} \\
\( T_{\mathrm{ref}} \) & Reference temperature & \(298\,\mathrm{K}\) & \cite{Chandesris2015MembraneDensity} \\
\( EW \) & Nafion equivalent weight & \(1.100\,\mathrm{kg/mol}\) & \cite{Sethuraman2008HydrogenCathode} \\
\( \rho_{\mathrm{Naf}} \) & Nafion density (dry) & \(1980\,\mathrm{kg/m}^3\) & \cite{Wong2014MacroscopicCells} \\
\( \varepsilon_{\mathrm{clc}} \) & Cathode layer thickness & \(1 \times 10^{-5}\,\mathrm{m}\) & \cite{Wong2014MacroscopicCells} \\
\( \gamma_{\mathrm{cat}} \) & Cathode rugosity & \(150\,\mathrm{m}^2/\mathrm{m}^2\) & \cite{Chandesris2015MembraneDensity} \\
\( k_1^0 \) & Base kinetic constant for H\(_2\)O\(_2\) formation & \(7.068 \times 10^{2}\,\mathrm{m}^7/\mathrm{mol}^2/\mathrm{s}\) & \cite{Sethuraman2008HydrogenCathode} \\
\( A_{\mathrm{H_2O_2}} \) & Activation energy for the ORR & \(42450\,\mathrm{J/mol}\) & \cite{Chandesris2015MembraneDensity} \\
\( \alpha_{\mathrm{H_2O_2}} \) & Transfer coefficient & \(0.5\) & \cite{Chandesris2015MembraneDensity} \\
\( \eta_{2e} \) & Cathodic Overpotential & \(0.695\) \, V & \cite{Chandesris2015MembraneDensity} \\

\bottomrule
\end{tabular}
\end{table}

\section{Free Radical Reactions}
\label{app:radical_reactions}

This appendix provides a detailed overview of the radical-driven reaction mechanisms considered in the degradation modeling of PEM water electrolyzers. These reactions involve highly reactive species such as hydroxyl (HO·) and hydroperoxyl (HOO·) radicals, which contribute to membrane chemical degradation. Unless otherwise stated, the reaction mechanisms and corresponding kinetic constants are compiled from the literature sources referenced in Section~\ref{sec:Mem_Modelling}. Table~\ref{tab:radical_reactions} summarizes the relevant reactions and their associated rate constants.
\begin{table}[h]
\scriptsize
\centering
\caption{Radical reactions and kinetic constants relevant to PEMWE degradation, based on \cite{Sethuraman2008HydrogenCathode} and \cite{Wong2014MacroscopicCells}.}
\label{tab:radical_reactions}
\renewcommand{\arraystretch}{1.2}
\begin{tabular}{|c|p{3.3cm}|p{3.7cm}|}
\hline
\textbf{\#} & \textbf{Reaction} & \textbf{Kinetic constant} \\
\hline
1 & \( \mathrm{O_2 + 2H^+ + 2e^- \rightarrow H_2O_2} \) & 
\( k_1 = k_1^0 \exp\left( \frac{-A_{\mathrm{H_2O_2}}}{RT} \right) \) \newline
\( \quad \cdot \exp\left( \frac{-\alpha_{\mathrm{H2O2}} F \eta_{2e}}{RT} \right) \) \newline
\( \quad [\mathrm{m^7/mol^2/s}] \) \\
\hline
2 & \( \mathrm{H_2O_2} \rightarrow 2\,\mathrm{HO^{\cdot}} \) & \( k_2 = 1.2 \times 10^{-7} \,[\mathrm{s^{-1}}] \) \\
\hline
3 & \( \mathrm{H_2O_2} + \mathrm{HO^{\cdot}} \rightarrow \mathrm{HOO^{\cdot}} + \mathrm{H_2O} \) & 
\( k_{3} = 2.7 \times 10^4 \,[\mathrm{m^3/mol/s}] \) \\
\hline
4 & \( \mathrm{O_2} + \mathrm{HO^{\cdot}} \rightarrow \mathrm{HOO^{\cdot}} + \mathrm{H_2O} \) & 
\( k_{4} = 1.2 \times 10^7 \,[\mathrm{m^3/mol/s}] \) \\
\hline
5 & \( \mathrm{HO^{\cdot}} + R_f{-}\mathrm{CF_2COOH} \rightarrow \text{products} + \mathrm{HF} \) & 
\( k_{5} \approx 10^3 \,[\mathrm{m^3/mol/s}] \) \\
\hline
\end{tabular}
\end{table}

\bibliography{references}     

\begin{thebibliography}{10}
\expandafter\ifx\csname url\endcsname\relax
  \def\url#1{\texttt{#1}}\fi
\expandafter\ifx\csname urlprefix\endcsname\relax\def\urlprefix{URL }\fi
\expandafter\ifx\csname href\endcsname\relax
  \def\href#1#2{#2} \def\path#1{#1}\fi

\bibitem{Barbir2005PEMSources}
F.~Barbir, \href{https://www.sciencedirect.com/science/article/pii/S0038092X04002464}{{PEM electrolysis for production of hydrogen from renewable energy sources}}, Solar Energy 78~(5) (2005) 661--669.
\newblock \href {https://doi.org/10.1016/J.SOLENER.2004.09.003} {\path{doi:10.1016/J.SOLENER.2004.09.003}}.
\newline\urlprefix\url{https://www.sciencedirect.com/science/article/pii/S0038092X04002464}

\bibitem{Sayed-Ahmed2024DynamicReview}
H.~Sayed-Ahmed, I.~Toldy, A.~Santasalo-Aarnio, \href{https://www.sciencedirect.com/science/article/pii/S1364032123007414}{{Dynamic operation of proton exchange membrane electrolyzers—Critical review}}, Renewable and Sustainable Energy Reviews 189 (2024) 113883.
\newblock \href {https://doi.org/10.1016/J.RSER.2023.113883} {\path{doi:10.1016/J.RSER.2023.113883}}.
\newline\urlprefix\url{https://www.sciencedirect.com/science/article/pii/S1364032123007414}

\bibitem{Benmehel2024PEMReflections}
A.~Benmehel, S.~Chabab, A.~L. Do~Nascimento~Rocha, M.~Chepy, T.~Kousksou, \href{https://www.sciencedirect.com/science/article/pii/S2590174524002162}{{PEM water electrolyzer modeling: Issues and reflections}}, Energy Conversion and Management: X 24 (2024) 100738.
\newblock \href {https://doi.org/10.1016/J.ECMX.2024.100738} {\path{doi:10.1016/J.ECMX.2024.100738}}.
\newline\urlprefix\url{https://www.sciencedirect.com/science/article/pii/S2590174524002162}

\bibitem{Pivac2024ReductionElectrolysis}
I.~Pivac, J.~{\v{S}}imunovi{\'{c}}, F.~Barbir, S.~Ni{\v{z}}eti{\'{c}}, \href{https://www.sciencedirect.com/science/article/abs/pii/S0360544224009307}{{Reduction of greenhouse gases emissions by use of hydrogen produced in a refinery by water electrolysis}}, Energy 296 (2024) 131157.
\newblock \href {https://doi.org/10.1016/J.ENERGY.2024.131157} {\path{doi:10.1016/J.ENERGY.2024.131157}}.
\newline\urlprefix\url{https://www.sciencedirect.com/science/article/abs/pii/S0360544224009307}

\bibitem{Liso2018ModellingTemperatures}
V.~Liso, G.~Savoia, S.~S. Araya, G.~Cinti, S.~K. K{\ae}r, {Modelling and experimental analysis of a polymer electrolyte membrane water electrolysis cell at different operating temperatures}, Energies 11~(12) (12 2018).
\newblock \href {https://doi.org/10.3390/en11123273} {\path{doi:10.3390/en11123273}}.

\bibitem{Schalenbach2013PressurizedCrossover}
M.~Schalenbach, M.~Carmo, D.~L. Fritz, J.~Mergel, D.~Stolten, \href{https://www.sciencedirect.com/science/article/pii/S0360319913022040}{{Pressurized PEM water electrolysis: Efficiency and gas crossover}}, International Journal of Hydrogen Energy 38~(35) (2013) 14921--14933.
\newblock \href {https://doi.org/10.1016/J.IJHYDENE.2013.09.013} {\path{doi:10.1016/J.IJHYDENE.2013.09.013}}.
\newline\urlprefix\url{https://www.sciencedirect.com/science/article/pii/S0360319913022040}

\bibitem{Falcao2020ABeginners}
D.~S. Falc{\~{a}}o, A.~M. Pinto, \href{https://www.sciencedirect.com/science/article/abs/pii/S0959652620312312}{{A review on PEM electrolyzer modelling: Guidelines for beginners}}, Journal of Cleaner Production 261 (2020) 121184.
\newblock \href {https://doi.org/10.1016/J.JCLEPRO.2020.121184} {\path{doi:10.1016/J.JCLEPRO.2020.121184}}.
\newline\urlprefix\url{https://www.sciencedirect.com/science/article/abs/pii/S0959652620312312}

\bibitem{Afshari2021PerformanceElectrolyzer}
E.~Afshari, S.~Khodabakhsh, N.~Jahantigh, S.~Toghyani, {Performance assessment of gas crossover phenomenon and water transport mechanism in high pressure PEM electrolyzer}, International Journal of Hydrogen Energy 46~(19) (2021) 11029--11040.
\newblock \href {https://doi.org/10.1016/J.IJHYDENE.2020.10.180} {\path{doi:10.1016/J.IJHYDENE.2020.10.180}}.

\bibitem{Li2024OptimizationPhenomenon}
Y.~Li, H.~Li, W.~Liu, Q.~Zhu, {Optimization of membrane thickness for proton exchange membrane electrolyzer considering hydrogen production efficiency and hydrogen permeation phenomenon}, Applied Energy 355 (2024) 122233.
\newblock \href {https://doi.org/10.1016/J.APENERGY.2023.122233} {\path{doi:10.1016/J.APENERGY.2023.122233}}.

\bibitem{Arunachalam2024EfficientApproach}
M.~Arunachalam, D.~S. Han, \href{https://link.springer.com/article/10.1007/s42247-024-00697-y}{{Efficient solar-powered PEM electrolysis for sustainable hydrogen production: an integrated approach}}, Emergent Materials 7~(4) (2024) 1401--1415.
\newblock \href {https://doi.org/10.1007/S42247-024-00697-Y/FIGURES/6} {\path{doi:10.1007/S42247-024-00697-Y/FIGURES/6}}.
\newline\urlprefix\url{https://link.springer.com/article/10.1007/s42247-024-00697-y}

\bibitem{Garcia-Valverde2012SimpleValidation}
R.~Garc{\'{i}}a-Valverde, N.~Espinosa, A.~Urbina, {Simple PEM water electrolyser model and experimental validation}, in: International Journal of Hydrogen Energy, Vol.~37, 2012, pp. 1927--1938.
\newblock \href {https://doi.org/10.1016/j.ijhydene.2011.09.027} {\path{doi:10.1016/j.ijhydene.2011.09.027}}.

\bibitem{Chandesris2015MembraneDensity}
M.~Chandesris, V.~M{\'{e}}deau, N.~Guillet, S.~Chelghoum, D.~Thoby, F.~Fouda-Onana, {Membrane degradation in PEM water electrolyzer: Numerical modeling and experimental evidence of the influence of temperature and current density}, International Journal of Hydrogen Energy 40~(3) (2015) 1353--1366.
\newblock \href {https://doi.org/10.1016/j.ijhydene.2014.11.111} {\path{doi:10.1016/j.ijhydene.2014.11.111}}.

\bibitem{Trinke2016HydrogenConditions}
P.~Trinke, B.~Bensmann, S.~Reichstein, R.~Hanke-Rauschenbach, K.~Sundmacher, \href{https://iopscience.iop.org/article/10.1149/2.0221611jes https://iopscience.iop.org/article/10.1149/2.0221611jes/meta}{{Hydrogen Permeation in PEM Electrolyzer Cells Operated at Asymmetric Pressure Conditions}}, Journal of The Electrochemical Society 163~(11) (2016) F3164--F3170.
\newblock \href {https://doi.org/10.1149/2.0221611JES/XML} {\path{doi:10.1149/2.0221611JES/XML}}.
\newline\urlprefix\url{https://iopscience.iop.org/article/10.1149/2.0221611jes https://iopscience.iop.org/article/10.1149/2.0221611jes/meta}

\bibitem{Frensch2019ImpactEmission}
S.~H. Frensch, G.~Serre, F.~Fouda-Onana, H.~C. Jensen, M.~L. Christensen, S.~S. Araya, S.~K. K{\ae}r, {Impact of iron and hydrogen peroxide on membrane degradation for polymer electrolyte membrane water electrolysis: Computational and experimental investigation on fluoride emission}, Journal of Power Sources 420 (2019) 54--62.
\newblock \href {https://doi.org/10.1016/j.jpowsour.2019.02.076} {\path{doi:10.1016/j.jpowsour.2019.02.076}}.

\bibitem{Sorace2022DevelopmentPhenomenon}
R.~Sorace, A.~Lanzini, \href{http://webthesis.biblio.polito.it/id/eprint/24955}{{Development and Analysis of Proton Exchange Membrane Water Electrolyzer Model with Chemical Degradation Phenomenon}}, Tech. rep., Politecnico di Torino (11 2022).
\newline\urlprefix\url{http://webthesis.biblio.polito.it/id/eprint/24955}

\bibitem{Feng2017AStrategies}
Q.~Feng, X.~Z. Yuan, G.~Liu, B.~Wei, Z.~Zhang, H.~Li, H.~Wang, \href{https://www.sciencedirect.com/science/article/abs/pii/S0378775317311631}{{A review of proton exchange membrane water electrolysis on degradation mechanisms and mitigation strategies}}, Journal of Power Sources 366 (2017) 33--55.
\newblock \href {https://doi.org/10.1016/J.JPOWSOUR.2017.09.006} {\path{doi:10.1016/J.JPOWSOUR.2017.09.006}}.
\newline\urlprefix\url{https://www.sciencedirect.com/science/article/abs/pii/S0378775317311631}

\bibitem{Siegmund2021CrossingElectrolysis}
D.~Siegmund, S.~Metz, V.~Peinecke, T.~E. Warner, C.~Cremers, A.~Grev{\'{e}}, T.~Smolinka, D.~Segets, U.~P. Apfel, \href{https://pmc.ncbi.nlm.nih.gov/articles/PMC8395688/}{{Crossing the Valley of Death: From Fundamental to Applied Research in Electrolysis}}, JACS Au 1~(5) (2021) 527.
\newblock \href {https://doi.org/10.1021/JACSAU.1C00092} {\path{doi:10.1021/JACSAU.1C00092}}.
\newline\urlprefix\url{https://pmc.ncbi.nlm.nih.gov/articles/PMC8395688/}

\bibitem{Bahr2020ArtificialStacks}
M.~Bahr, A.~Gusak, S.~Stypka, B.~Oberschachtsiek, \href{/doi/pdf/10.1002/cite.202000089 https://onlinelibrary.wiley.com/doi/full/10.1002/cite.202000089}{{Artificial Neural Networks for Aging Simulation of Electrolysis Stacks}}, Chemie-Ingenieur-Technik 92~(10) (2020) 1610--1617.
\newblock \href {https://doi.org/10.1002/CITE.202000089;PAGEGROUP:STRING:PUBLICATION} {\path{doi:10.1002/CITE.202000089;PAGEGROUP:STRING:PUBLICATION}}.
\newline\urlprefix\url{/doi/pdf/10.1002/cite.202000089 https://onlinelibrary.wiley.com/doi/full/10.1002/cite.202000089}

\bibitem{Hayatzadeh2024MachineCatalyst}
A.~Hayatzadeh, M.~Fattahi, A.~Rezaveisi, \href{https://www.sciencedirect.com/science/article/abs/pii/S0360319923064492}{{Machine learning algorithms for operating parameters predictions in proton exchange membrane water electrolyzers: Anode side catalyst}}, International Journal of Hydrogen Energy 56 (2024) 302--314.
\newblock \href {https://doi.org/10.1016/J.IJHYDENE.2023.12.149} {\path{doi:10.1016/J.IJHYDENE.2023.12.149}}.
\newline\urlprefix\url{https://www.sciencedirect.com/science/article/abs/pii/S0360319923064492}

\bibitem{Courtois2023CanDomingos}
A.~Courtois, J.~M. Morel, P.~Arias, \href{https://link.springer.com/article/10.1007/s13398-023-01411-z}{{Can neural networks extrapolate? Discussion of a theorem by Pedro Domingos}}, Revista de la Real Academia de Ciencias Exactas, Fisicas y Naturales - Serie A: Matematicas 117~(2) (2023) 1--26.
\newblock \href {https://doi.org/10.1007/S13398-023-01411-Z/METRICS} {\path{doi:10.1007/S13398-023-01411-Z/METRICS}}.
\newline\urlprefix\url{https://link.springer.com/article/10.1007/s13398-023-01411-z}

\bibitem{Raissi2019Physics-informedEquations}
M.~Raissi, P.~Perdikaris, G.~E. Karniadakis, {Physics-informed neural networks: A deep learning framework for solving forward and inverse problems involving nonlinear partial differential equations}, Journal of Computational Physics 378 (2019) 686--707.
\newblock \href {https://doi.org/10.1016/J.JCP.2018.10.045} {\path{doi:10.1016/J.JCP.2018.10.045}}.

\bibitem{Hu2024Physics-informedApplications}
H.~Hu, L.~Qi, X.~Chao, \href{https://www.sciencedirect.com/science/article/pii/S0263823124009364}{{Physics-informed Neural Networks (PINN) for computational solid mechanics: Numerical frameworks and applications}}, Thin-Walled Structures 205 (2024) 112495.
\newblock \href {https://doi.org/10.1016/J.TWS.2024.112495} {\path{doi:10.1016/J.TWS.2024.112495}}.
\newline\urlprefix\url{https://www.sciencedirect.com/science/article/pii/S0263823124009364}

\bibitem{Moradi2023NovelPDEs}
S.~Moradi, B.~Duran, S.~Eftekhar~Azam, M.~Mofid, \href{https://www.mdpi.com/2075-5309/13/3/650/htm https://www.mdpi.com/2075-5309/13/3/650}{{Novel Physics-Informed Artificial Neural Network Architectures for System and Input Identification of Structural Dynamics PDEs}}, Buildings 2023, Vol. 13, Page 650 13~(3) (2023) 650.
\newblock \href {https://doi.org/10.3390/BUILDINGS13030650} {\path{doi:10.3390/BUILDINGS13030650}}.
\newline\urlprefix\url{https://www.mdpi.com/2075-5309/13/3/650/htm https://www.mdpi.com/2075-5309/13/3/650}

\bibitem{Chen2024Physics-InformedGeoengineering}
X.~X. Chen, P.~Zhang, Z.~Y. Yin, \href{https://www.tandfonline.com/doi/pdf/10.1080/17499518.2024.2315301}{{Physics-Informed neural network solver for numerical analysis in geoengineering}}, Georisk: Assessment and Management of Risk for Engineered Systems and Geohazards 18~(1) (2024) 33--51.
\newblock \href {https://doi.org/10.1080/17499518.2024.2315301} {\path{doi:10.1080/17499518.2024.2315301}}.
\newline\urlprefix\url{https://www.tandfonline.com/doi/pdf/10.1080/17499518.2024.2315301}

\bibitem{Mishra2022EstimatesPDEs}
S.~Mishra, R.~Molinaro, \href{https://dx.doi.org/10.1093/imanum/drab032}{{Estimates on the generalization error of physics-informed neural networks for approximating a class of inverse problems for PDEs}}, IMA Journal of Numerical Analysis 42~(2) (2022) 981--1022.
\newblock \href {https://doi.org/10.1093/IMANUM/DRAB032} {\path{doi:10.1093/IMANUM/DRAB032}}.
\newline\urlprefix\url{https://dx.doi.org/10.1093/imanum/drab032}

\bibitem{Bensmann2022AnDevelopment}
B.~Bensmann, A.~Rex, R.~Hanke-Rauschenbach, {An engineering perspective on the future role of modelling in proton exchange membrane water electrolysis development} (6 2022).
\newblock \href {https://doi.org/10.1016/j.coche.2022.100829} {\path{doi:10.1016/j.coche.2022.100829}}.

\bibitem{Zerrougui2025Physics-InformedElectrolysis}
I.~Zerrougui, Z.~Li, D.~Hissel, \href{https://www.sciencedirect.com/science/article/pii/S2666546825000060}{{Physics-Informed Neural Network for modeling and predicting temperature fluctuations in proton exchange membrane electrolysis}}, Energy and AI 20 (2025) 100474.
\newblock \href {https://doi.org/10.1016/J.EGYAI.2025.100474} {\path{doi:10.1016/J.EGYAI.2025.100474}}.
\newline\urlprefix\url{https://www.sciencedirect.com/science/article/pii/S2666546825000060}

\bibitem{Chen2024MachineFramework}
X.~Chen, A.~Rex, J.~Woelke, C.~Eckert, B.~Bensmann, R.~Hanke-Rauschenbach, P.~Geyer, {Machine learning in proton exchange membrane water electrolysis — A knowledge-integrated framework}, Applied Energy 371 (10 2024).
\newblock \href {https://doi.org/10.1016/j.apenergy.2024.123550} {\path{doi:10.1016/j.apenergy.2024.123550}}.

\bibitem{Marangio2009TheoreticalProduction}
F.~Marangio, M.~Santarelli, M.~Cal{\`{i}}, \href{https://www.sciencedirect.com/science/article/pii/S0360319908016571}{{Theoretical model and experimental analysis of a high pressure PEM water electrolyser for hydrogen production}}, International Journal of Hydrogen Energy 34~(3) (2009) 1143--1158.
\newblock \href {https://doi.org/10.1016/J.IJHYDENE.2008.11.083} {\path{doi:10.1016/J.IJHYDENE.2008.11.083}}.
\newline\urlprefix\url{https://www.sciencedirect.com/science/article/pii/S0360319908016571}

\bibitem{Biaku2008AElectrolyzer}
C.~Y. Biaku, N.~V. Dale, M.~D. Mann, H.~Salehfar, A.~J. Peters, T.~Han, \href{https://www.sciencedirect.com/science/article/pii/S0360319908007118}{{A semiempirical study of the temperature dependence of the anode charge transfer coefficient of a 6 kW PEM electrolyzer}}, International Journal of Hydrogen Energy 33~(16) (2008) 4247--4254.
\newblock \href {https://doi.org/10.1016/J.IJHYDENE.2008.06.006} {\path{doi:10.1016/J.IJHYDENE.2008.06.006}}.
\newline\urlprefix\url{https://www.sciencedirect.com/science/article/pii/S0360319908007118}

\bibitem{Carmo2013AElectrolysis}
M.~Carmo, D.~L. Fritz, J.~Mergel, D.~Stolten, \href{https://www.sciencedirect.com/science/article/abs/pii/S0360319913002607}{{A comprehensive review on PEM water electrolysis}}, International Journal of Hydrogen Energy 38~(12) (2013) 4901--4934.
\newblock \href {https://doi.org/10.1016/J.IJHYDENE.2013.01.151} {\path{doi:10.1016/J.IJHYDENE.2013.01.151}}.
\newline\urlprefix\url{https://www.sciencedirect.com/science/article/abs/pii/S0360319913002607}

\bibitem{Wallnofer-Ogris2024AResearch}
E.~Walln{\"{o}}fer-Ogris, I.~Grimmer, M.~Ranz, M.~H{\"{o}}glinger, S.~Kartusch, J.~Rauh, M.~G. Macherhammer, B.~Grabner, A.~Trattner, \href{https://www.sciencedirect.com/science/article/pii/S0360319924012692}{{A review on understanding and identifying degradation mechanisms in PEM water electrolysis cells: Insights for stack application, development, and research}}, International Journal of Hydrogen Energy 65 (2024) 381--397.
\newblock \href {https://doi.org/10.1016/J.IJHYDENE.2024.04.017} {\path{doi:10.1016/J.IJHYDENE.2024.04.017}}.
\newline\urlprefix\url{https://www.sciencedirect.com/science/article/pii/S0360319924012692}

\bibitem{Kim2021AccurateMembranes}
S.~H. Kim, B.~T.~D. Nguyen, H.~Ko, M.~Kim, K.~Kim, S.~Y. Nam, J.~F. Kim, \href{https://www.sciencedirect.com/science/article/abs/pii/S0360319921005243#fig3}{{Accurate evaluation of hydrogen crossover in water electrolysis systems for wetted membranes}}, International Journal of Hydrogen Energy 46~(29) (2021) 15135--15144.
\newblock \href {https://doi.org/10.1016/J.IJHYDENE.2021.02.040} {\path{doi:10.1016/J.IJHYDENE.2021.02.040}}.
\newline\urlprefix\url{https://www.sciencedirect.com/science/article/abs/pii/S0360319921005243#fig3}

\bibitem{PatrickTrinke2021ExperimentalElectrolyzers}
H.~M.~S. Patrick~Trinke, \href{https://www.repo.uni-hannover.de/handle/123456789/11061}{{Experimental and model-based investigations on gas crossover in polymer electrolyte membrane water electrolyzers}} (2021).
\newline\urlprefix\url{https://www.repo.uni-hannover.de/handle/123456789/11061}

\bibitem{Ruvinskiy2011UsingReaction}
P.~S. Ruvinskiy, A.~Bonnefont, C.~Pham-Huu, E.~R. Savinova, \href{/doi/pdf/10.1021/la2006343}{{Using ordered carbon nanomaterials for shedding light on the mechanism of the cathodic oxygen reduction reaction}}, Langmuir 27~(14) (2011) 9018--9027.
\newblock \href {https://doi.org/10.1021/LA2006343/SUPPL{\_}FILE/LA2006343{\_}SI{\_}001.PDF} {\path{doi:10.1021/LA2006343/SUPPL{\_}FILE/LA2006343{\_}SI{\_}001.PDF}}.
\newline\urlprefix\url{/doi/pdf/10.1021/la2006343}

\bibitem{Wong2014MacroscopicCells}
K.~H. Wong, E.~Kjeang, \href{https://iopscience.iop.org/article/10.1149/2.0031409jes https://iopscience.iop.org/article/10.1149/2.0031409jes/meta}{{Macroscopic In-Situ Modeling of Chemical Membrane Degradation in Polymer Electrolyte Fuel Cells}}, Journal of The Electrochemical Society 161~(9) (2014) F823--F832.
\newblock \href {https://doi.org/10.1149/2.0031409JES/XML} {\path{doi:10.1149/2.0031409JES/XML}}.
\newline\urlprefix\url{https://iopscience.iop.org/article/10.1149/2.0031409jes https://iopscience.iop.org/article/10.1149/2.0031409jes/meta}

\bibitem{Davini2021UsingNetworks}
D.~Davini, B.~Samineni, B.~Thomas, A.~H. Tran, C.~Zhu, K.~Ha, G.~Dasika, L.~White, {Using physics-informed regularization to improve extrapolation capabilities of neural networks}, Fourth workshop on machine learning and the physical sciences (NeurIPS 2021) (2021).

\bibitem{Cai2021Physics-informedReview}
S.~Cai, Z.~Mao, Z.~Wang, M.~Yin, G.~E. Karniadakis, \href{https://link.springer.com/article/10.1007/s10409-021-01148-1}{{Physics-informed neural networks (PINNs) for fluid mechanics: a review}}, Acta Mechanica Sinica/Lixue Xuebao 37~(12) (2021) 1727--1738.
\newblock \href {https://doi.org/10.1007/S10409-021-01148-1/METRICS} {\path{doi:10.1007/S10409-021-01148-1/METRICS}}.
\newline\urlprefix\url{https://link.springer.com/article/10.1007/s10409-021-01148-1}

\bibitem{Jalili2024Physics-informedFlows}
D.~Jalili, S.~Jang, M.~Jadidi, G.~Giustini, A.~Keshmiri, Y.~Mahmoudi, \href{https://www.sciencedirect.com/science/article/pii/S0017931023012346}{{Physics-informed neural networks for heat transfer prediction in two-phase flows}}, International Journal of Heat and Mass Transfer 221 (2024) 125089.
\newblock \href {https://doi.org/10.1016/J.IJHEATMASSTRANSFER.2023.125089} {\path{doi:10.1016/J.IJHEATMASSTRANSFER.2023.125089}}.
\newline\urlprefix\url{https://www.sciencedirect.com/science/article/pii/S0017931023012346}

\bibitem{Mayyas2019ManufacturingElectrolyzers}
A.~T. Mayyas, M.~F. Ruth, B.~S. Pivovar, G.~Bender, K.~B. Wipke, \href{https://www.osti.gov/servlets/purl/1557965/}{{Manufacturing Cost Analysis for Proton Exchange Membrane Water Electrolyzers}} (8 2019).
\newblock \href {https://doi.org/10.2172/1557965} {\path{doi:10.2172/1557965}}.
\newline\urlprefix\url{https://www.osti.gov/servlets/purl/1557965/}

\bibitem{Colbertaldo2017Zero-dimensionalElectrolyzers}
P.~Colbertaldo, S.~L. G{\'{o}}mez~Al{\'{a}}ez, S.~Campanari, \href{https://www.sciencedirect.com/science/article/pii/S1876610217363506}{{Zero-dimensional dynamic modeling of PEM electrolyzers}}, Energy Procedia 142 (2017) 1468--1473.
\newblock \href {https://doi.org/10.1016/J.EGYPRO.2017.12.594} {\path{doi:10.1016/J.EGYPRO.2017.12.594}}.
\newline\urlprefix\url{https://www.sciencedirect.com/science/article/pii/S1876610217363506}

\bibitem{Awasthi2011DynamicProduction}
A.~Awasthi, K.~Scott, S.~Basu, \href{https://www.sciencedirect.com/science/article/pii/S0360319911006343}{{Dynamic modeling and simulation of a proton exchange membrane electrolyzer for hydrogen production}}, International Journal of Hydrogen Energy 36~(22) (2011) 14779--14786.
\newblock \href {https://doi.org/10.1016/J.IJHYDENE.2011.03.045} {\path{doi:10.1016/J.IJHYDENE.2011.03.045}}.
\newline\urlprefix\url{https://www.sciencedirect.com/science/article/pii/S0360319911006343}

\bibitem{Sethuraman2008HydrogenCathode}
V.~A. Sethuraman, J.~W. Weidner, A.~T. Haug, S.~Motupally, L.~V. Protsailo, {Hydrogen Peroxide Formation Rates in a PEMFC Anode and Cathode}, Journal of The Electrochemical Society 155~(1) (2008) B50.
\newblock \href {https://doi.org/10.1149/1.2801980} {\path{doi:10.1149/1.2801980}}.

\end{thebibliography}

\end{document}